\documentclass[]{elsart}
 \usepackage{graphicx}

\journal{Physics Letters B}

\usepackage{amssymb}
\usepackage{epsfig}

\newcommand{\beq}{\begin{equation}}
\newcommand{\eeq}{\end{equation}}
\newcommand{\bea}{\begin{eqnarray}}
\newcommand{\eea}{\end{eqnarray}}
\newcommand{\ave}[1]{\langle {#1} \rangle}

\newcommand{\dslash}{\partial\!\!\!/}

\newcommand{\pb}{\bar\psi}

\newcommand{\eq}[1]{Eq.~(\ref{#1})}

\def\roughly#1{\mathrel{\raise.3ex\hbox{$#1$\kern-.75em%
\lower1ex\hbox{$\sim$}}}}

\def\={\;=\;}
\def\+{\;+\;}

\def\eps{\varepsilon}

\def\bear{\begin{array}}
\def\ear{\end{array}}

\def\newline{\hfil\break}

\begin{document}
\begin{frontmatter}
\title{Neutron stars and the transition to superconducting quark matter}

\author[label1]{M. Baldo},
\author[label2,label3]{M. Buballa\corauthref{cor}}\ead{michael.buballa@physik.tu-darmstadt.de},
\author[label1]{G.F. Burgio},
\author[label2]{F. Neumann}, 
\author[label4]{M. Oertel}, and
\author[label1]{H.-J. Schulze}
\address[label1]{INFN, Sezione di Catania, 57 Corso Italia, 
95129 Catania, Italy}
\address[label2]{Institut f\"ur Kernphysik, TU Darmstadt,
Schlossgartenstr. 9, 64289 Darmstadt, Germany}
\address[label3]{Gesellschaft f\"ur Schwerionenforschung,
Planckstr. 1, 64291 Darmstadt, Germany}
\address[label4] {IPN-Lyon, 43 Bd du 11 Novembre 1918,
69622 Villeurbanne C\'edex, France}
\corauth[cor]{Corresponding author}
\begin{abstract}

We explore the relevance of color superconductivity inside 
a possible quark matter core for the bulk properties of neutron
stars.  For the quark phase we use a Nambu--Jona-Lasinio (NJL)
type model, extended to include diquark condensates.  
For the hadronic phase, a microscopic many-body model
is adopted, with and without strangeness content.
In our calculations, a sharp boundary is assumed between the hadronic 
and the quark phases. 
For NJL model parameters fitted to vacuum properties
we find that no star with a pure quark core does exist. 
Nevertheless the presence of color superconducting phases can lower the 
neutron star maximum mass substantially.
In some cases, the transition to quark matter occurs only if color 
superconductivity is present. Once the quark phase is introduced,
the value of the maximum mass stays in any case below 
the value of two solar masses.  

\end{abstract}
\begin{keyword}
dense matter \sep equation of state \sep hadron-quark phase transition \sep 
neutron stars \sep color superconductivity


\PACS 26.60.+c \sep 24.10.Cn  \sep 12.39.-x  \sep 25.75.Nq 

\end{keyword}

\end{frontmatter}
\section{Introduction}
\label{1}
According to QCD, the fundamental theory of strong interaction, 
nuclear matter is viewed as the confined phase of quark-gluon matter,
where chiral symmetry is broken and quarks are bound inside nucleons.
At large enough density, nuclear matter is expected to undergo a phase
transition to the deconfined phase, where quarks and gluons are free to
move in the medium and chiral symmetry is restored. Unfortunately,
this transition is not well understood, since QCD lattice
calculations cannot yet be performed at large density, i.e., at large
chemical potential. Experimentally, in ultra-relativistic heavy ion
collisions the transition to the deconfined phase is expected to occur 
at high temperature but essentially at zero baryon density. On the other hand,
neutron stars are believed to contain very high baryon density in their
interior, where the transition to the deconfined phase could occur.
It has been argued by several authors that the bulk properties of
neutron stars can be strongly affected by the presence of a core
where a quark phase or a mixed hadron-quark phase is present.
The fact that the quark phase - if present - is likely to be
a color superconductor has recently attracted much 
attention (For reviews see \cite{RaWi00,Alford01} and references
therein). In general, the condensation energy associated with the
superconductivity is only a small fraction of the total energy,
and the bulk properties of the equation of state (EOS) are hardly affected.
This is because only a small fraction of fermions around the Fermi
surface participates effectively to pairing, and the condensation energy is 
not proportional to the pairing gap $\Delta$, but to $\Delta^2/E_F$,
where $E_F$ is the Fermi energy. These considerations equally apply
to color superconductivity in quark matter. However, as recently
argued~\cite{AlRe02}, the structure
of neutron stars appears to be sensitive to the properties of the
possible quark phase and even a relatively small change in the
quark equation of state could affect the predictions of neutron star
properties. 
\par 
In this paper we study the consequence of
introducing explicitly color superconducting phases in the deconfined 
quark matter for the structure and bulk properties of neutron stars. 
To this purpose, we describe the hadronic phase by the EOS 
derived from a microscopic many-body theory \cite{bbb}. For the quark matter
we adopt a three-flavor Nambu-Jona Lasinio (NJL) type model, which has been
recently extended to include diquark condensates~\cite{BO02,SRP02,NBO02}.
In contrast to bag models which are often employed to describe 
quark phases, the NJL model dynamically generates a density-dependent
effective bag constant and density dependent effective quark masses
which can be considerably larger than the bare masses.
It has been shown earlier that this has important consequences
for the (non-) existence of absolutely stable strange quark matter
\cite{BO99} and the color-flavor (un-) locking phase transition in
quark matter~\cite{BO02}. In the context of neutron star interiors
the NJL model has been employed in Refs.~\cite{SLS99} and \cite{SPL00}. 
It was found by both groups that quarks exist at most in a small mixed 
phase regime but not in a pure quark core. This could be traced back 
to the relatively large values for effective bag constant and the effective 
strange quark mass which results from the calculation.   
On the other hand these calculations did not include diquark condensates.
As recently discussed within a bag model~\cite{AlRe02} the latter effectively
lower the bag constant and could have sizable consequences 
for the hadron-quark phase transition and for the properties of compact stars.
It is therefore interesting to study these effects for an NJL model 
equation of state.

Depending on the flavors which participate in a diquark
condensate one can distinguish between several color superconducting
quark phases, most important the two-flavor superconducting (2SC)
phase~\cite{ARW98,RSSV98} 
where only up and down quarks are paired and the color-flavor
locked (CFL) phase~\cite{ARW99} 
which contains $ud$, $us$, and $ds$ pairs.  In
principle this can give rise to a large number of globally electric
and color neutral mixed quark phases which are, however, unlikely to
be stable if Coulomb and surface effects are included~\cite{NBO02}.
Therefore in this letter we discuss only homogeneous neutral quark
phases which have been constructed first in Ref.~\cite{SRP02}. We have
checked, however, that the use of the mixed quark phase equation of
state of Ref.~\cite{NBO02} yields practically the same results.

Similarly,
we assume a sharp (first order) transition from the hadronic to the
quark phase.  This is motivated by the results of
Ref.~\cite{ARRW01} where it was found that a quark-hadron mixed phase
is unlikely to be stable for reasonable values of the surface tension.
We will briefly comment on the possible consequences of a mixed phase in 
Sec.~\ref{stars}. 

\section{Hadronic and quark EOS}
\par\noindent
We  start  with the description of the hadronic phase. 
The EOS is based on
the Brueckner--Bethe--Goldstone (BBG) many-body theory, which is 
a linked cluster 
expansion of the energy per nucleon of nuclear matter (see Ref.~\cite{book},
chapter 1 and references therein).  
It has been shown
that the non-relativistic BBG expansion is well convergent  \cite{song,BaBu01},
and  the Brueckner-Hartree-Fock (BHF) level of approximation is accurate
in the density range relevant for neutron  stars.  
The basic ingredient in this many--body approach is the Brueckner reaction 
matrix $G$, which is the solution of the  Bethe--Goldstone equation 

\begin{equation}
G[n;\omega] = v  + \sum_{k_a k_b} v {{|k_a k_b\rangle  Q  \langle k_a k_b|}
  \over {\omega - e(k_a) - e(k_b) }} G[n;\omega], 
\end{equation}                                                           
\noindent
where $v$ is the bare nucleon-nucleon (NN) interaction, $n$ is the nucleon 
number density, and $\omega$ the  starting energy.  
The single-particle energy $e(k)$ (assuming $\hbar$=1 here and throughout 
the paper),
\begin{equation}
e(k) =  {{k^2}\over {2m}} + U(k),
\label{e:en}
\end{equation}
\noindent
and the Pauli operator $Q$ determine the propagation of intermediate 
baryon pairs. The Brueckner--Hartree--Fock (BHF) approximation for the 
single-particle potential
$U(k)$  using the  {\it continuous choice} prescription is
\begin{equation}
U(k) = {\rm Re} \sum _{k'\leq k_F} \langle k k'|G[n; e(k)+e(k')]|k k'\rangle_a,
\end{equation}
\noindent
where the subscript ``{\it a}'' indicates antisymmetrization of the 
matrix element.  
Due to the occurrence of $U(k)$ in Eq.~(\ref{e:en}), they constitute 
a coupled system that has to be solved in a self-consistent manner
for several Fermi momenta of the particles involved. 
In the BHF approximation the energy per nucleon is
\begin{equation}
{E \over{A}}  =  
          {{3}\over{5}}{{k_F^2}\over {2m}}  + {{1}\over{2n}}  
~ {\rm Re} \sum_{k,k'\leq k_F} \langle k k'|G[n; e(k)+e(k')]|k k'\rangle_a. 
\end{equation}
\noindent
In the calculations reported here we have used the Paris potential
\cite{lac80} as the two-nucleon interaction and the Urbana model as
three-body force \cite{CPW83,schi}.  
The corresponding nuclear matter EOS fulfills several requirements,
namely (i) it reproduces the correct nuclear matter saturation point
$\rho_0$ \cite{bbb},
(ii) the incompressibility is compatible with the values extracted
from phenomenology, (iii) the symmetry energy is compatible with
nuclear phenomenology, (iv) the causality condition is always
fulfilled.  \par Recently, we have included the hyperon degrees of
freedom within the same approximation to calculate the nuclear EOS
needed to describe the NS interior \cite{bbs}. We have included the
$\Sigma^-$ and $\Lambda$ hyperons. To this purpose, one needs also
nucleon-hyperon  and hyperon-hyperon  interactions
\cite{bbs,barc}.  However, because of a lack of experimental data, the
hyperon-hyperon interaction has been neglected in the first
approximation in this work, whereas for the nucleon-hyperon interaction the
Nijmegen soft-core model \cite{mae89} has been adopted.  
\par 
In neutron stars one has to consider matter in beta equilibrium,
where electrons and eventually muons coexist with baryons,
while neutrinos are considered to escape from the star.
The EOS for the beta equilibrated matter can be obtained 
once the hadron matter is known, together with the chemical
potentials of different species as a function of total baryon 
density. Since the procedure is standard, we do not give further details 
of the calculations.
\par
For the quark phase we consider a leptonic contribution
from electrons and muons and a quark contribution. The former is
treated in free gas approximation. For the latter we employ the model
defined by the Lagrangian
\beq 
\mathcal{ L}_{eff} \= \bar{\psi} (i \dslash
- \hat{m}) \psi \+ \mathcal{L}_{q\bar q} \+ \mathcal{L}_{qq},
\label{Lagrange}
\end{equation}
where $\psi$ denotes a quark field with three flavors and three colors. 
The mass matrix $\hat m$ has the form 
$\hat m = {\rm diag}(m_u, m_d, m_s)$ in flavor space.
We consider an NJL-type interaction with a quark-quark part
\beq
    \mathcal{L}_{qq} \=
    H\sum_{A = 2,5,7} \sum_{A' = 2,5,7}
    (\pb \,i\gamma_5 \tau_A \lambda_{A'} \,C\pb^T)
    (\psi^T C \,i\gamma_5 \tau_A \lambda_{A'} \, \psi) 
    \,
\label{Lqq}
\eeq
and a quark-antiquark part
\bea
    \mathcal{L}_{q\bar q} &\=& G\, \sum_{a=0}^8 \Big[(\pb \tau_a\psi)^2
    \+ (\pb i\gamma_5 \tau_a \psi)^2\Big]\nonumber \\ && 
    \;-\; K\,\Big[{\rm det}_f\Big(\pb(1+\gamma_5)\psi\Big) \,+\
                   {\rm det}_f\Big(\pb(1-\gamma_5)\psi\Big)\Big]\;.
\label{Lqbarq}
\eea
Here $\tau_i$ and $\lambda_j$ are $SU(3)$- (Gell-Mann-)matrices in
flavor and color space, respectively.  $\tau_0 =
\sqrt{\frac{2}{3}}\,1\hspace{-1.5mm}1_f$ is proportional to the unit
matrix in flavor space
\footnote{In this form the interaction is 
understood to be used at mean-field level in Hartree approximation.}.
In order to determine the energy density and the pressure of the
system we calculate the mean-field (grand canonical)
thermodynamic potential $\Omega(T=0,\{\mu_i\})$ in the presence of
three quark-antiquark condensates $\ave{\bar u u}$, $\ave{\bar d d}$,
and $\ave{\bar s s}$, and three possible diquark condensates
corresponding to $ud$, $us$, and $ds$ pairing in the scalar color-
antitriplet channel. The different chemical potentials $\{\mu_i\}$
are associated with the conserved charges of the system and
are adjusted such that we deal with color and electrically neutral
quark matter in beta equilibrium. 
More details about the model can be found in
Refs.~\cite{BO02,SRP02,NBO02}, where it has been employed to study
color superconducting quark matter. 

Since the model is not renormalizable we have to specify a
regularization procedure. For simplicity we will use a
three-dimensional momentum cutoff $\Lambda$. The values of $\Lambda =
602.3$~MeV, the two coupling constants $G = 1.835/\Lambda^2$ and $K
= 12.36/\Lambda^5$ as well as the current quark masses $m_u = m_d =
5.5$~MeV, $m_s = 140.7$~MeV are taken from Ref.~\cite{Rehberg}, where
they have been adjusted to reproduce masses and decay constants of
the pseudoscalar meson nonet. For the quark-quark coupling constant we take
$H=G$, as motivated in Ref.~\cite{NBO02}. This is likely to be an 
upper limit of the realistic values for this coupling constant.  

We should mention that we neglect possible contributions of the
``pseudo''-Goldstone bosons of broken chiral symmetry to $\Omega$ in
the CFL phase.  Since in neutral CFL matter the electric
charge chemical potential vanishes~\cite{SRP02,NBO02} no
charged bosons are excited. The contributions of a possible
$K_0$-condensate~\cite{BS02} are expected to be small 
(see, e.g., the discussion in
Ref.~\cite{AlRe02}) and will therefore be neglected.

\section{Hadron-quark phase transition}
\label{phasetr}

\begin{figure}[t]
\parbox{13cm}{\hspace{2cm}
     \epsfig{file=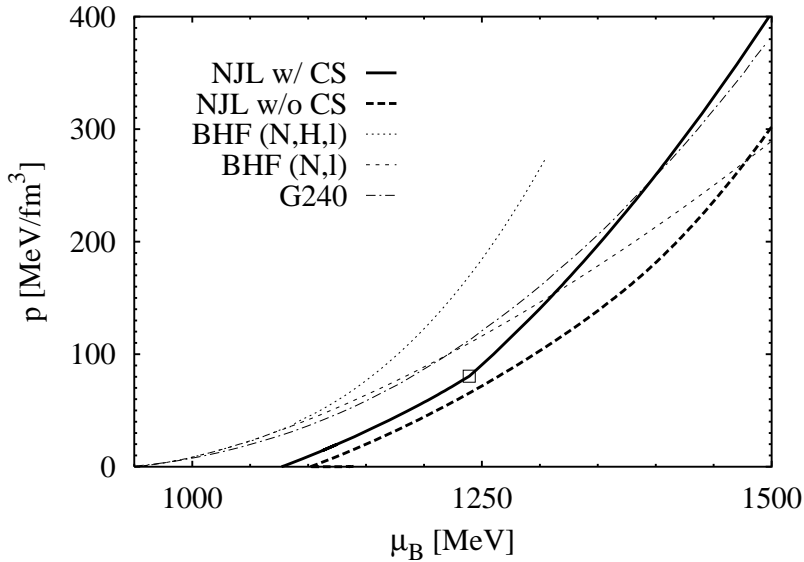, width=11.cm}}
\caption{\it Pressure of neutral hadron and quark matter in
beta equilibrium as a function
of baryon chemical potential: BHF
calculation without hyperons (dashed) and with hyperons
(dotted), relativistic mean field EOS with 
$K = 240$~MeV~\cite{Glendenning}
(dashed-dotted), and NJL quark
matter (bold lines) without (dashed) and with CS (solid). The open square
indicates the transition point from the 2SC phase to the CFL phase.}
\label{figp}
\end{figure}

In Fig.~\ref{figp} the pressure of neutral hadron and quark matter in
beta equilibrium is displayed as a function of the baryon chemical potential.
For the hadronic case, we consider the microscopic EOS including 
only nucleons and leptons as well as also including hyperons
(dashed and dotted lines, respectively).
The presence of strange matter considerably softens \cite{bbs} the
EOS, which results in a steeper increase of pressure as a function of
baryon chemical potential.\par
The pressure of quark matter with and 
without color superconductivity (CS) is indicated by the bold lines.
In both EOS strangeness is included.
(For CS in neutral NJL quark matter without strangeness
see \cite{huang}).  
The EOS without CS (bold dashed line) is identical to the one employed
in Refs.~\cite{BO99,SLS99}. Here the strangeness content is zero below
$\mu_B = 1295$~MeV and rises smoothly above this point.
This is quite different for the case with color superconductivity
(solid line). Strictly speaking, this line corresponds to two phases,
separated by a first order phase transition at $\mu_0 \simeq 1240$~MeV:
the 2SC phase for $\mu_B < \mu_0$ and the CFL phase for $\mu_B > \mu_0$.
At $\mu_B = \mu_0$ (
open square in the figure)
the number density of strange quarks increases discontinuously 
from almost zero in the 2SC phase to 1/3 of the total quark number
density in the CFL phase. Thereby the total baryon density
jumps from less than four to more than five times nuclear matter 
density. This corresponds to a sudden increase of the slope of the curve,
which is clearly visible in the figure.
\par
Starting from the surface of a neutron star, the baryon density and the 
baryon chemical potential increase
as one moves inside, until a possible transition
to quark matter occurs. In the considered plot of pressure vs.  
baryon chemical potential, the transition is marked by the crossing
of the hadron EOS with the quark matter EOS. 
If only nucleons are included in the hadronic sector, this crossing
indeed occurs. However, the presence of CS substantially lowers
the density at which the transition happens (see Table~\ref{tabletrans}). 
Obviously, this is closely related to the kink at the 2SC-CFL 
transition point: 
In the 2SC phase the introduction of color superconductivity leads only
to a moderate enhancement $\delta p$ of the pressure which can be attributed 
to some additional binding caused by the formation of Cooper pairs.
However, in the CFL phase $\delta p$ grows faster and thereby reduces
the critical chemical potential for the hadron-quark phase transition 
by about 165~MeV. As discussed above, this increased slope corresponds
to a higher density which is mainly due to the sudden appearance
of a large amount of strange quarks in the CFL phase. 
\begin{table}[t]
\begin{center}
\begin{tabular}{|l|c c c|}
\hline
transition & BHF(N,l) $\rightarrow$ w/o CS & 
BHF(N,l) $\rightarrow$ CFL & G240 $\rightarrow$ CFL
\\
\hline
$\mu_B$ (MeV)         & 1478 & 1312 & 1397 \\
$\rho_B^{(h)}/\rho_0$ &  4.6 &  3.7 &  6.5 \\
$\rho_B^{(q)}/\rho_0$ &  7.9 &  6.4 &  7.7 \\
$\Delta$ (MeV)        &  --- &  115 &  120 \\
$M_u$ (MeV)           &   10 &   28 &   22 \\
$M_s$ (MeV)           &  283 &  261 &  236 \\
$B_{\mathit{eff}}$ (${\rm MeV}/{\rm fm}^3$) & 176 & 208 & 234 \\
\hline
\end{tabular}
\end{center}
\caption{\small\it Various quantities at the phase transition points from 
         hadronic matter to NJL quark matter(see Figs.~\ref{figp} and 
         \ref{figeos}): chemical potential, baryon density in the hadronic 
         phase ($\rho_B^{(h)}$) and in the quark phase ($\rho_B^{(q)}$)
         in units of normal nuclear matter density 
         $\rho_0 = 0.17\,{\rm fm}^{-3}$, average diquark gap 
         $\Delta = \sqrt{(\Delta_{ud}^2 + \Delta_{us}^2 + \Delta_{ds}^2)/3}$, 
         effective quark masses, and effective bag constant.}
\label{tabletrans}
\end{table}  

The microscopic hadronic EOS which includes hyperons 
does not present any crossing with the quark EOS up to large density,
even if color superconducting phases are included.
In that case there would be no transition to a quark phase at densities
relevant for neutron stars. However,
it has already been found in Ref. \cite{bbs} that this EOS gives a maximum
neutron star mass of
about 1.25 solar masses (see next section), which is below the
observational limit of 1.44 solar masses~\cite{hulse}. Therefore, this EOS 
cannot describe the whole interior of neutron stars, and
either this particular EOS or the NJL description of the quark phase 
must be ruled out.
But even if we keep the considered quark EOS,
the appearance of hyperons cannot be excluded, since the microscopic 
EOS depends on the adopted hyperon-hyperon interaction,
at least at high density. In fact, while the hyperon-nucleon interaction
is constrained, to some extent, by  hyperon-nucleus
phenomenology, very little is known about the hyperon-hyperon interaction
potential. It is therefore possible that the hyperon-hyperon interaction
becomes strongly repulsive at higher density, and the corresponding
EOS is actually stiffer than the adopted one. To illustrate
the relevance of this uncertainty, we have considered a phenomenological
hadronic EOS based on relativistic mean field scheme of 
Ref.~\cite{Glendenning} with $K = 240$~MeV (hereafter called `G240').
Without hyperons
this EOS turns out to be very similar to our microscopic EOS.
Once hyperons are introduced, the 
microscopic (dotted) and mean field (dashed-dotted) 
EOS are quite different above the hyperon threshold, 
which indicates that the hyperon interactions are quite different
in the two cases. With the mean field EOS, a phase transition is now possible,
but only if CS is included in the quark EOS.
Due to this uncertainty on the hyperon interactions, it is not possible
to firmly establish if and in which place the quark phase can 
appear once hyperons are considered. As we will see, this uncertainty 
will not affect our main conclusions. 

\begin{figure}[t]
\parbox{13cm}{\hspace{2cm}
     \epsfig{file=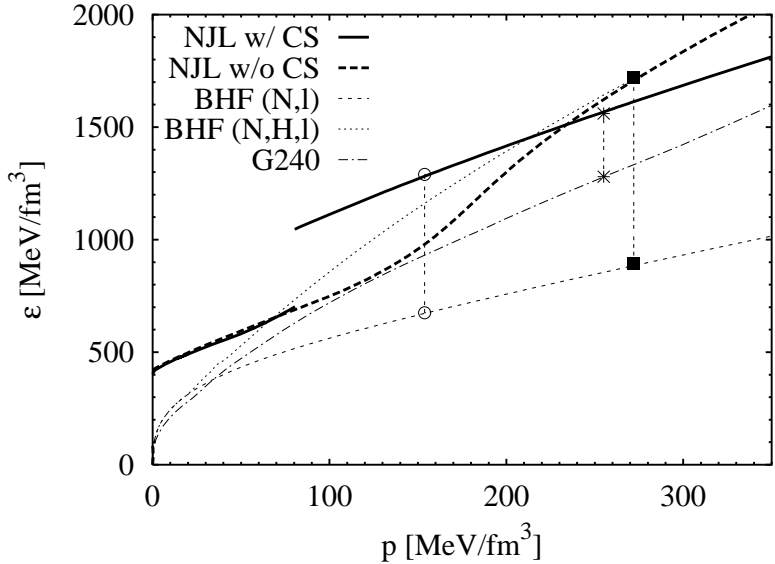, width=11.cm}}
\caption{\it Hadronic and quark matter equations of state. The various
lines have the same meaning as in Fig.~\ref{figp}.
The points indicate the pressure where the respective hadron-quark
phase transitions occur.}
\label{figeos}
\end{figure}
\par
In Fig.~\ref{figeos} the energy density as a function of pressure
is displayed for the various EOS discussed above. 
We have also indicated the points where the hadron-quark phase transitions 
take place, as obtained from Fig.~\ref{figp}. The open circles and the full 
squares mark the phase transition from the microscopic EOS (with nucleons
only) to quark matter with and without CS, respectively. 
The asterisks correspond to the transition from the mean field EOS
(with hyperons) to the quark EOS with CS.

As discussed above, the quark EOS with CS (solid line) 
contains a first order phase transition from the 2SC phase to the CFL phase.
At the transition point ($p \simeq 80$~MeV/fm$^3$),
this leads to a strong discontinuity in the energy density.
As a consequence, the effect of including color superconducting phases
in the NJL model is qualitatively different from the behavior in a 
bag model, like in Ref.~\cite{AlRe02}. In a bag model, the
formation of Cooper pairs more or less acts like a negative contribution to
the bag constant, reducing the energy density and enhancing the
pressure for a given chemical potential. 
In our case the situation is more complicated. If we compare the solid line
with the bold dashed line we see that color superconductivity {\it enhances}
the energy density in a certain regime of pressure. The reason is again
the fact that the CFL phase always contains a large amount of (relatively
heavy!) strange quarks which strongly contribute to the energy density,
whereas the NJL quark matter without CS contains no or very few
strange quarks up to much larger values of $p$.

For our later discussion it is useful to introduce 
effective bag model parameters to characterize the NJL EOS.
To that end one inserts the effective quark masses
and diquark gaps of the NJL model into the bag model expression for the 
pressure and defines an effective bag constant $B_{\mathit{eff}}$
by equating the result with the pressure obtained in the NJL model.
Note, however, that in contrast to a bag model $B_{\mathit{eff}}$,
the gaps, and effective quark masses are density dependent 
quantities. In Table~\ref{tabletrans} they are given for the
chemical potentials corresponding to the phase transition points 
from the hadronic phase.
Compared with most bag models, both, the strange quark mass and the
bag constant, are relatively large. As we will see below, this has
important consequences for the structure of compact stars. 

\section{Neutron star structure}
\label{stars}

\begin{figure}[t]
\parbox{11cm}{\hspace{2cm}\epsfig{file=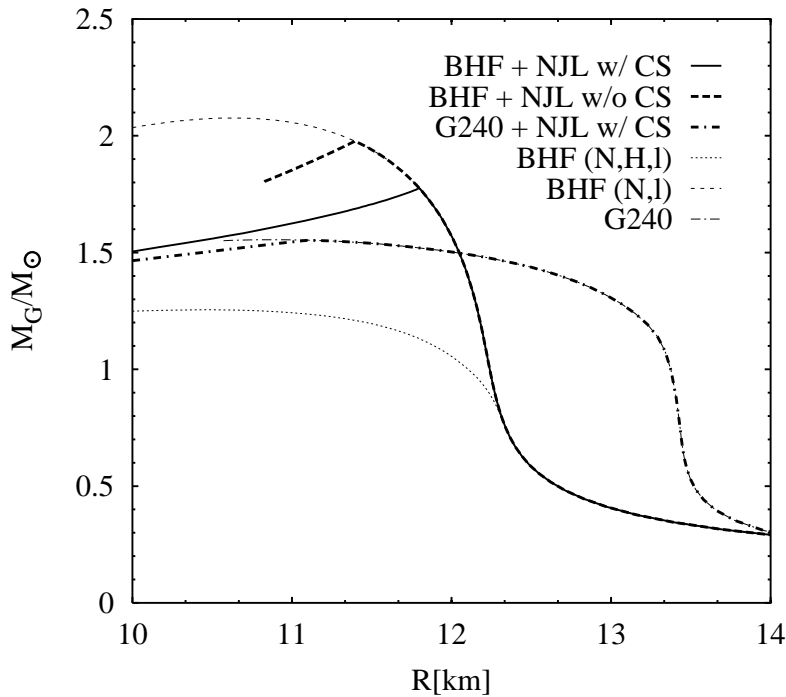, width=9.cm, angle=-90}}
\caption{\it Gravitational masses of compact stars as functions of the 
radius for the different EOS discussed in Table~\ref{tablestars}.}
\label{figmass}
\end{figure}

For a given EOS the mass of a static compact star as a function of its
radius can be obtained by solving the Tolman-Oppenheimer-Volkoff equation.
The resulting curves for the EOS constructed above are displayed in  
Fig.~\ref{figmass}. The corresponding maximum masses and maximum
central energy densities are listed in Table~\ref{tablestars}.
We are particularly interested in the possible existence of hybrid stars
with a pure quark core. According to the results of Sec.~\ref{phasetr} 
there are three cases where such a hybrid star could be expected, 
namely for the microscopic EOS without hyperons combined with a quark EOS 
with or without CS (solid and bold dashed line, respectively) and for the
relativistic mean field EOS combined with the quark EOS with CS 
(bold dashed-dotted line).
As a consequence of the discontinuous energy density at the transition point
(see Fig.~\ref{figeos}) the phase transitions manifest themselves by cusps in 
the mass-radius relation.  
It turns out, however, that in all three cases this cusp is strong enough
to render the star unstable. 
Hence, if our EOS are correct, no star with a pure quark core can exist. 

This result seems to be rather insensitive to the choice of the hadronic
EOS and must mainly be attributed to the quark EOS derived within the
NJL model.
Recently, Alford and Reddy have found stable hybrid stars with pure
quark cores described within a bag model EOS~\cite{AlRe02}.
The hadronic EOS employed in that analysis are comparable to
our microscopic EOS without hyperons. 
For a bag constant $B = 137$~MeV/fm$^3$ and a strange quark mass
$M_s = 200$~MeV the authors found that a star with a pure quark core is 
unstable without CS, but stable if a (CFL) diquark gap of 100~MeV is
chosen. 
If we compare the above numbers with the effective quantities
listed in Table~\ref{tabletrans}, we see that the NJL model is
characterized by relatively large values of the strange quark mass
and the effective bag constant. This leads to considerably larger
energy densities in the quark phase which are finally responsible
for the instability.

Of course, the existence of a mixed phase, instead of a sharp transition,
would smooth out the cusps in Fig.~\ref{figmass} and a quark component
could be present in the core of the neutron star. However, the value 
of the maximum mass is not expected to be modified by a substantial
amount. 

\begin{table}[t]
\begin{center}
\begin{tabular}{|l|c|c|}
\hline
EOS  &  $M_G^{max}/M_\odot$  & $\eps_c^{max}$ (MeV/fm$^3$)  \\ \hline 
BHF(N,l) & 2.07 & 1409 \\
BHF(N,l) + NJL without CS & 1.97 & \phantom{1}884 \\
BHF(N,l) + NJL with CS  & 1.77 & \phantom{1}672 \\
BHF(N,H,l) & 1.25 & 1442 \\
G240 & 1.55 & 1434 \\
G240 + NJL with CS & 1.55 & 1279  \\
\hline
\end{tabular}
\end{center}
\caption{\small\it Maximum gravitational mass $M_G^{max}$ and corresponding
                central energy density $\eps_c^{max}$ for the various
                EOS.}
\label{tablestars}
\end{table}  

\section{Conclusions} 
 
We have combined various hadronic EOS with an NJL model EOS with and
without color superconductivity to construct the hadron-quark phase
transition and to investigate the structure of compact stars.
Our results do not allow for a pure quark phase in the interior
of a neutron star. In some cases we find no phase transition at all
while in other cases the star becomes unstable as soon as the phase
transition occurs. Both phenomena are mainly due to the relatively
large values of the effective strange quark mass and the effective
bag constant in the NJL model.
Considering non-superconducting quark matter we thus confirm the
results of Ref.~\cite{SLS99}. Including color superconductivity
does not change these findings.  

Of course, the properties of the NJL EOS depend on model
parameters. In this letter we have adopted the parameters of 
Ref.~\cite{Rehberg} which have been adjusted to vacuum properties.
Although this is common practice in this field, it is very well
possible that these parameters are not appropriate to describe
quark matter at densities relevant for compact stars. 
Thus our arguments could also be turned around:  
If there were strong hints for the existence of pure quark cores
in compact stars, this would indicate a considerable modification
of the effective NJL-type quark interactions in dense 
matter~\footnote{After submission of our manuscript, two groups have reported
the construction of a stable hybrid star with a pure quark matter core
described within two-flavor NJL-type models including color superconductivity,
but without strange quarks.
This is in strong contradiction to our results where even a phase
transition to quark matter seems to be impossible without strange
quarks.
In Ref.~\cite{SHH03} this was achieved within a standard NJL model
with relatively light up and down quarks, corresponding to a relatively 
small effective bag constant.
The authors of Ref.~\cite{BGAYT} find stable quark cores if the
integrals are regularized by Gaussian form factors, but not for a
sharp cut-off.
In both papers the hadronic phase is described within relativistic
mean field models.
These findings call for a systematic survey of the parameter dependence 
of the results and of the influence of possible extensions and modifications
of the model.}.

If there are no stable quark cores in compact stars,
color superconductivity is unlikely to exist in any natural surrounding. 
Nevertheless, the possibility
of CS could have an effect on the maximum mass
of neutron stars, as is evident from the difference between the
solid line and the bold dashed line in Fig.~\ref{figmass}.
While {\it excluding} CS the
maximum mass can be close to a value of two solar masses, 
{\it including} CS the maximum
mass is reduced to 1.77 solar masses in our
calculations. (If we had chosen a smaller value
for the quark-quark coupling constant $H$ in \eq{Lqq} the maximum
mass would of course come out somewhere in between these two values.)
Introducing hyperons softens
the EOS further, and the corresponding maximum mass is further lowered. 
If the
softening is large, as, e.g., in the case of the EOS of
Ref.~\cite{bbs}, the neutron star becomes unstable already before 
the transition density to a possible quark matter core is
reached. In this case the maximum mass falls below the observational limit.  
\par 
In conclusion, the inclusion of CS in the
quark phase keeps the neutron star maximum mass well below two solar
masses, independently of the details of the hadronic EOS. 
The observation of a neutron star with a mass well above two solar
masses would seriously question our present view on the EOS of
quark matter and on the structure of the quark matter phase.
\\[2mm]
{\bf Acknowledgments:}\\
M.B. thanks INFN
for financial support during his visit to Catania in November 2002
and Marcello Baldo and Fiorella Burgio for their warm hospitality.
M.O. acknowledges support from the Alexander von
Humboldt foundation as a Feodor-Lynen fellow.

\end{document}